\documentclass[aps,prl,reprint,preprintnumbers,showpacs,showkeys,superscriptaddress]{revtex4-1}
\usepackage{amsmath,amssymb,bm,mathrsfs,graphicx}
\usepackage[colorlinks=true,citecolor=blue,linkcolor=blue]{hyperref}

\begin{document}
\title{Englert--Brout--Higgs Mechanism in Nonrelativistic Systems}

\author{Haruki Watanabe}
\email{hwatanabe@berkeley.edu}
\affiliation{Department of Physics, University of California, Berkeley, California 94720, USA}

\author{Hitoshi Murayama}
\email{hitoshi@berkeley.edu, hitoshi.murayama@ipmu.jp}
\affiliation{Department of Physics, University of California, Berkeley, California 94720, USA} 
\affiliation{Theoretical Physics Group, Lawrence Berkeley National Laboratory, Berkeley, California 94720, USA}
\affiliation{Kavli Institute for the Physics and Mathematics of the Universe (WPI), Todai Institutes for Advanced Study, University of Tokyo, Kashiwa 277-8583, Japan} 

\begin{abstract}
We study the general theory of Englert--Brout--Higgs mechanism without assuming Lorentz invariance.
In the presence of a finite expectation value of non-Abelian matter charges, gauging those symmetries always results in spontaneous breaking of spatial rotation.  
If we impose the charge neutrality by assuming a
background with the opposite charges, the dynamics of the background
cannot be decoupled and has to be fully taken into account.
In either case, the spectrum is continuous as the gauge coupling is switched off.
\end{abstract}

\preprint{IPMU14-0109, UCB-PTH-14/30}
\maketitle

\paragraph{Introduction.}---The discovery of the Higgs boson at the
Large Hadron Collider marks a great triumph of unity of physics.  The
original idea emerged from the study of superconductivity and its
theory by Bardeen, Cooper, and Schrieffer (BCS)~\cite{Bardeen:1957mv}.
After Anderson found that there are collective excitations in the gap
region~\cite{PhysRev.110.827}, Nambu first clarified that the BCS
ground state is still consistent with gauge invariance~\cite{PhysRev.117.648}, and introduced the concept of spontaneous
symmetry breaking (SSB) into particle physics~\cite{PhysRev.122.345,PhysRev.124.246}.  Soon afterwards, Goldstone
proved that the SSB leads to massless scalar particles called
Nambu--Goldstone bosons (NGBs)~\cite{Goldstone:1961eq}.  Even though
the original theorem did not apply to non-Lorentz-invariant systems,
the present authors generalized the theorem so that it has real-life
applications in condensed matter physics, atomic physics, nuclear
physics, and astrophysics~\cite{Hagen,WatanabeMurayama1,Hidaka}.  On the other
hand, Englert, Brout~\cite{Englert:1964et}, and Higgs~\cite{Higgs:1964pj} proposed the gauged symmetry with SSB to go around
Goldstone's theorem because Nature does not appear to have a massless
scalar boson.  It is called Englert--Brout--Higgs (EBH) mechanism.  
The Higgs boson predicted by this mechanism was finally discovered fifty years later.  At the same time, the
concept of the Higgs boson made a full circle back to condensed matter
physics, becoming a hot subject of research (see, {\it e.g.}\/, Refs.~\cite{Varma2002,PhysRevB.86.054508}). 

Given this tremendous cross-pollination among different subareas in
physics, it is natural to ask what the general theory of Higgs
phenomenon is without relying on the Lorentz invariance.  In
particular, the non-abelian gauge theory developed in particle physics~\cite{Yang:1954ek} is making inroads into condensed matter physics,
such as spin liquid~\cite{Wen}, multi-layer graphene~\cite{PhysRevLett.108.216802,PhysRevB.86.081403}, ultracold atoms in
optical lattices~\cite{Tagliacozzo,PhysRevLett.110.125303,PhysRevLett.110.125304} etc.
Therefore a specific question of great importance is: {\it what is the
general theory of the EBH mechanism in non-abelian gauge theories
without Lorentz invariance?}\/

In relativistic field theories, spontaneously broken gauge symmetries
do not give rise to physical NGBs --- would-be NGBs are {\it eaten} by
gauge fields and in turn gauge fields acquire the longitudinal
component and a finite mass.  This is how the EBH mechanism works.  It
turns out that the extension of this famous story to the general
non-relativistic setup is not a trivial problem.  In particular, the
emergence of type-A and type-B NGBs~\cite{WatanabeMurayama1,Hidaka}
not seen in Lorentz-invariant
systems raises questions about the number of ``eaten'' degrees of
freedom, as there are only half as many type-B NGBs as the broken
symmetries.

In pioneering works~\cite{Shovkovy1,Shovkovy2}, Gusynin and his
collaborators studied a gauged $\text{U}(2)$ linear $\sigma$ model
with a finite chemical potential.  The chemical potential breaks the
Lorentz symmetry, allowing the matter field to have a finite
expectation value of non-Abelian charge densities $\langle
j_{I(\text{matter})}^{\mu=0}\rangle\neq0$ in the ground state.  (In this Letter, we use $I,J$ for generators of internal symmetries.) They found that the gauge field develops a finite expectation value $\langle \vec{A}^J\rangle$ and spontaneously
breaks the spatial rotation.  However, it has been left unclear
whether the spontaneous breaking of spatial rotation is required by a
physical principle or just a peculiar property of the specific model.

On the other hand, Hama and his collaborators~\cite{Hama} discussed
the EBH mechanism for the same model but in a physically distinct
setup; namely, they assumed the presence of a background that
neutralizes the total matter charge densities of the system, {\it i.e.,} $\langle
j_{(\text{matter})}^{\mu=0}\rangle+\langle j_{\text{(bg)}}^{\mu=0}\rangle=0$.  In their analysis, they simply subtracted the
background contribution in the form $eA_\mu^I\langle
j_{I\text{(bg)}}^{\mu}\rangle$ from the Lagrangian, generalizing the
prescription called ``charge neutrality'' for Abelian cases discussed in
Refs.~\cite{Kapusta,Kapusta2} to non-Abelian charge densities.  They
then found that the system does not show spontaneous breaking of
spatial rotation.  However, the excitation spectrum in their analysis
appears discontinuous as a function of gauge coupling
$e$~\footnote
{For example, see Fig.~2 of Ref.~\cite{Hama}.  The mode with a gap 
$2\mu$ at $e=0$ is missing in the limit $e\rightarrow0$, since the
mass gap of the gauge bosons goes to zero in their analysis.}.  At
least for a weak coupling and perturbative physics, which is the main focus of this Letter, the low-energy
spectrum must be continuous as a function of the gauge coupling.

Given these previous studies, in this Letter, we show the following
statements in order. (i) When we do not impose the neutrality of
non-Abelian matter charges (and hence with type-B
  NGBs) and if we gauge the corresponding non-Abelian
symmetries, the system {\it must}\/ break the spatial
rotation in the weak coupling regime.  (ii) When we assume the presence of a background that
neutralizes the net matter charges of the system, the dynamics of
the background {\it cannot}\/ be decoupled and we have to include it
explicitly as a dynamical degrees of freedom; otherwise the symmetry
of the system is explicitly violated, ending up with the unphysical
discontinuity of the spectrum in Ref.~\cite{Hama}.  

Throughout the Letter, we exclusively consider the three spatial
dimensions $d=3$ to avoid the discussions of the Chern-Simons terms.

\paragraph{Necessity of Spontaneous Breaking of Spatial Rotation.}
---The Higgs mechanism becomes nontrivial when at least one commutator
of Noether charges acquires a nonzero expectation values at $e=0$.
The minimal symmetry breaking pattern that realizes this behavior is
$\text{SO}(3)\rightarrow \text{SO}(2)$: a ferromagnet has a nonzero
expectation value of $Q_3=-i[Q_1,Q_2]$, while it vanishes for
antiferromagnet~\footnote{We consider only the continuous part of the
  symmetry group and do not discuss, {\it e.g.}\/, the time reversal symmetry.}.  Therefore, here we compare the gauged version of the ferromagnet and antiferromagnet to illustrate our point in the simplest example.

Let us start with the gauged antiferromagnet. The low-energy effective Lagrangian at $e=0$ reads $\mathcal{L}=\mathcal{L}_{\text{(matter)}}+\mathcal{L}_{\text{(gauge)}}$, where
\begin{eqnarray}
\mathcal{L}_{\text{(matter)}}&=&(\rho/2v^2)\dot{\vec{n}}\cdot\dot{\vec{n}}-(\rho/2)\partial_i\vec{n}\cdot\partial_i\vec{n},\\
\mathcal{L}_{\text{(gauge)}}&=&\text{tr}F_{ti}F_{ti}-(c^2/2)\text{tr}F_{ij}F_{ij}.\label{gauge1}
\end{eqnarray}
Here, $\vec{n}$ is a unit vector, $A_\mu=A_\mu^IT_I=A_\mu^I\sigma_I/2$ ($\sigma_{I=1,2,3}$ are Pauli matrices) is the $\text{SO}(3)$ gauge field, and $F_{\mu\nu}=\partial_\mu A_\nu-\partial_\nu A_\mu+ie[A_\mu,A_\nu]$ is the field strength. $c$ is the speed of light in the medium and $i=x,y,z$ is the spatial index. $\mathcal{L}_{\text{(matter)}}$ describes two (type-A) NGBs with the dispersion relation $\omega=vk$, while $\mathcal{L}_{\text{(gauge)}}$ describes $3(d-1)$ photons with $\omega=ck$.  

At a finite gauge coupling $e\neq0$~\footnote{For the effective Lagrangian of these models at $e\neq0$ before taking the unitary gauge, see Ref.~\cite{SM}.}, one may take the unitary gauge to set $\vec{n}=(0,0,1)$ and the Lagrangian becomes
\begin{eqnarray}
\mathcal{L}=(\rho/2v^2)e^2A_t^aA_t^a-(\rho/2)e^2\vec{A}^a\cdot\vec{A}^a+\mathcal{L}_{\text{(gauge)}}.\label{antiferrounitarygauge}
\end{eqnarray}
Each broken gauge field $A_\mu^{a=1,2}$ acquires both the mass gap and the longitudinal component. Transverse components describe $(d-1)$ gapped modes with $\omega_T=\sqrt{(ck)^2+e^2\rho}$, while the longitudinal (and the temporal) component produce(s) a gapped mode with $\omega_L=\sqrt{(vk)^2+e^2\rho}$.  The unbroken gauge field $A_\mu^z$ remains unchanged.  Note that the gap of $\omega_T$ and $\omega_L$ are the same as dictated by the $\text{SO}(d)$ spatial rotation.  In other words, $\vec{A}^a$ at $\vec{k}=0$ describes a massive spin-one vector boson, who has $d$ states. In the limit $e\rightarrow0$, they all become gapless: the transverse components reproduce photons and the longitudinal components reproduce each type-A NGB.  This situation is exactly the same as the Lorentz invariant case.  For this to be the case, in general, the number of NGBs (associated with gauged symmetries) at $e=0$ and the number of massive vector bosons at $e>0$ must be the same.

Now we turn to the gauged ferromagnet. The matter part of the effective Lagrangian $\mathcal{L}_{\text{(matter)}}$ at $e=0$ reads 
\begin{eqnarray}
m_3(n_2\dot{n}_1-n_1\dot{n}_2)/(1+n_3)-(\rho/2)(\partial_i\vec{n})^2\label{ferromagnetmatter}
\end{eqnarray}
and $\mathcal{L}_{\text{(gauge)}}$ is the same as Eq.~\eqref{gauge1}.  Here, $m_3\equiv\langle Q_3\rangle/V=-i\langle [Q_1,Q_2]\rangle/V$ represents the magnetization.  This case, $\mathcal{L}_{\text{(matter)}}$ describes only one (type-B) NGB with a quadratic dispersion $\omega=\rho k^2/m_3$.  For $e\neq0$, the Lagrangian in the unitary gauge reads
\begin{eqnarray}
\mathcal{L}=-em_3A_{\mu=t}^{I=3}-(\rho/2)e^2\vec{A}^a\cdot\vec{A}^a+\mathcal{L}_{\text{(gauge)}}.\label{ferrounitarygauge}
\end{eqnarray}
We can already see at this point the necessity of spontaneous breaking
of the {\it spatial} rotation.  If the $\text{SO}(d)$ spatial rotation
were not broken, each broken gauge field $A_\mu^{a}$ at $\vec{k}=0$
would be a massive vector boson with $d$ states.  Hence, in the limit
$e\rightarrow0$, they would produce $2d$ gapless modes. However, there
are only $2(d-1)+1$ gapless modes (two photons and one type-B NGB) at $e=0$. This is a contradiction.

Our argument here did not use any details of the example.  Whenever a charge commutator acquires a vacuum expectation value, the number of NGBs becomes less than the number of broken symmetries.  When these symmetries are gauged, one encounters the same difficulty and the spontaneous breaking of the spatial $\text{SO}(d)$ rotation is inevitable.

\paragraph{Charge Neutrality by Gauge Fields.}
---In order to see what is going on more concretely, let us examine the Yang--Mills equation $D_\mu F^{\mu\nu}=ej_{\text{(matter)}}^\nu$, or equivalently,
\begin{eqnarray}
\partial_\mu F^{\mu\nu}&=&ej_{\text{(total)}}^\nu\equiv ej_{\text{(matter)}}^\nu+ej_{\text{(gauge)}}^\nu.\label{YM}
\end{eqnarray}
Here, $j_{\text{(total)}}^\nu$ is the conserved Noether current for
the global part of the gauge symmetry, and
\begin{eqnarray}
j_{\text{(matter)}}^\mu&\equiv&-T_Ie^{-1}\partial\mathcal{L}_\text{(matter)}/\partial A_\mu^I,\\
j_{\text{(gauge)}}^\mu&\equiv&-T_Ie^{-1}\partial\mathcal{L}_\text{(gauge)}/\partial A_\mu^I=i[A_\nu,F^{\mu\nu}].\label{gaugecurernt}
\end{eqnarray}
Especially, for a homogeneous field configuration $\partial_\mu=0$, Eq.~\eqref{YM} is reduced to
\begin{eqnarray}
j_{\text{(total)}}^\mu=j_{\text{(matter)}}^\mu+j_{\text{(gauge)}}^\mu=0,
\end{eqnarray}
meaning that, even if we start with $\langle
j_{\text{(matter)}}^0\rangle\neq0$ at $e=0$, the gauge field $A_\mu$
condenses to cancel it and the total current
$j_{\text{(total)}}^\mu$ vanishes for any finite $e$.  There might be a solution with $\langle j_{\text{(matter)}}^0\rangle=\langle j_{\text{(gauge)}}^0\rangle=0$, but this solution is clearly discontinuous at $e=0$ if $\langle j_{\text{(matter)}}^0\rangle|_{e=0}\neq0$.  Instead, continuous solutions should take a value $\langle j_{\text{(gauge)}}^0\rangle=-\langle j_{\text{(matter)}}^0\rangle|_{e=0}+O(e)\neq0$ for a small enough $e$.

The temporal component of the gauge current~\eqref{gaugecurernt} for a homogeneous solution is given by
\begin{equation}
j_{\text{(gauge)}}^0=e[A_i,[A_0,A_i]].\label{gaugecurrent2}
\end{equation}
Thus, in the weak coupling regime, both $A^0$ and $\vec{A}$ should condense to realize $\langle j_{\text{(gauge)}}^0\rangle\neq0$, spontaneously breaking the spatial rotation by $\langle\vec{A}\rangle$.  For example, in the case of the ferromagnet, we have
\begin{equation}
\langle A_{\mu=t}^{I=3}\rangle=\sqrt{\rho},\quad \langle A_{\mu=z}^{I=1}\rangle=-\sqrt{m_3/(e\sqrt{\rho})}.
\end{equation}
The spatial $\text{SO}(d)$ rotation is spontaneously broken to $\text{SO}(d-1)$ rotation around the $z$ axis and the internal $\text{SO}(d)$ symmetry is completely broken.  

The expectation value of a spatial component of the gauge field is
reminiscent of the Hosotani mechanism
\cite{Hosotani:1983xw,Csaki:2002ur}, where gauge fields in compactified
dimensions play the role of the eaten NGBs.  This also occurs when
boundary conditions violate the gauge symmetry \cite{Csaki:2003dt}.

To obtain the excitation spectrum, we expand the Lagrangian around this expectation value to the quadratic order in fluctuation.  The calculation is particularly easy when we set the momentum along the $z$ axis $\vec{k}=k\hat{z}$.  As we show step by step in our Supplemental Material~\cite{SM}, six fields $A_{\mu=z,0}^{I=1,2,3}$ produce only one physical mode, whose dispersion is given by
\begin{equation}
\omega=\sqrt{\left(\rho k^2/m_3\right)^2-3e\sqrt{\rho}\,(\rho k^2/m_3)+4e^2\rho}.
\end{equation}
This clearly becomes the type-B NGB in the limit $e\rightarrow 0$.
Three fields $A_{i}^{I=1,2,3} (i\neq z)$ produce three physical modes with gaps at $k=0$:
\begin{equation}
0,\quad \kappa c,\quad\sqrt{(\kappa c)^2+4e^2\rho},\label{spectrumferromagnet}
\end{equation}
where we defined $\kappa\equiv-e \langle A_{\mu=z}^{I=1}\rangle=\sqrt{em_3/\sqrt{\rho}}>0$.  The first mode is one of the two (type-A) NGBs corresponding to the spontaneously broken spatial rotation.   All of these three modes reproduce photons $\omega=ck$ in the limit $e\rightarrow 0$.

Due to the unbroken $\text{SO}(d-1)$ rotation, $d-1$ components $A_{i}^{I=1,2,3} (i\neq z)$ give the same spectrum. However, 
thanks to the broken $\text{SO}(d)$ rotation, $A_{\mu=z}^{I=1,2,3}$ does not produce the same spectrum as in~\eqref{spectrumferromagnet} even at $k=0$.

Unfortunately, this is not the end of the story for this particular
example.  If we take a closer look at the first mode in
Eq.~\eqref{spectrumferromagnet}, one notices that the energy has zero
not only at $k=0$ but also at a finite momentum $k=\kappa$.  Moreover,
even when one sets the momentum along $x$-axis $\vec{k}=k\hat{x}$, one
finds a similar zero at $k=\kappa$ in a mode. This implies a possibility of formation of a three-dimensional crystalline order, which we have not worked out yet and should be definitely an interesting future work.  Note however that our claim of spontaneous breaking of spatial rotation remains true even in this case.

Turning back to the general case, let us make three remarks in order.  

First, our discussion above assumed the continuity at
$e=0$. Indeed, it was confirmed in Ref.~\cite{Gongyo}
  that the number of degrees of freedom is continuous, even though
  they did not study the spectrum.  But it is of course possible in principle that there is no perturbative regime and physics is discontinuous as a function of the gauge coupling constant.

Second, for a bigger non-Abelian symmetry group, the existence of homogenous solutions of the Yang--Mills equation $e\langle[A_\nu,[A^\mu,A^\nu]]\rangle=\langle j_{\text{(matter)}}^\mu\rangle$ is quite nontrivial.  Take $\text{SU}(3)$ as an example and assume $\langle j_{\text{(matter)}}^\mu\rangle|_{e=0}=\delta^{\mu 0}m_8 T_8\neq0$. ($T_I\equiv\lambda_I/2$ and $\lambda_{I=1,\ldots,8}$ are Gell-Mann matrices).  In this case, it is not sufficient to condense only one of $\vec{A}^I$. Instead, for instance, $A_{\mu=z}^{I=5}\neq0$ and $A_{\mu=y}^{I=7}\neq0$ works.  This is because there is no root vector of $\text{SU}(3)$ algebra along $T_8$. Note that condensing enough number of $A_i^I$ in different directions is only possible in sufficiently large spatial dimensions. Note also that there may not be any unbroken rotation.

Third, we can of course understand the EBH mechanism in any gauges.  Let us in particular discuss the $R_{\xi}$ gauge again using the ferromagnet.  As explained above, the internal $\text{SO}(3)$ symmetry is completely broken and there are no net charge densities in the ground state $\langle j_{\text{(total)}}^0\rangle=0$. Therefore, from the general counting rule established in Ref.~\cite{WatanabeMurayama1}, we expect three would-be type-A NGBs. This number coincides with the number of broken gauge symmetries.  We regard $A_{\mu=1,\ldots,d-1}^{I}$ as vector fields of the unbroken $\text{SO}(d-1)$ rotation and $A_{\mu=0,d}^{I}$ and $\vec{n}$ as scaler fields.  In the $R_{\xi}$ gauge, one introduces gauge fixing terms to eliminate the linear coupling between vectors and scalers.  One can then reproduce the above spectrum and also establish the non-relativistic version of the equivalence theorem.

\paragraph{Neutralizing Charges by a Background.}---The whole nontrivial consequence of the EBH mechanism in the above discussion originated from a finite non-Abelian matter charge density $\langle j_{\text{(matter)}}^0\rangle|_{e=0}\neq0$.  Here we consider the situation where $\langle j_{\text{(matter)}}^0\rangle|_{e=0}$ is cancelled by the opposite contribution from a background.

Let us first review the Abelian case.  Our example is a superconductor,
which exhibits the EBH mechanism via the condensation of Cooper pairs
of electrons.  Clearly there is a $\text{U}(1)$ charge density of electrons $e\langle j_{\text{(matter)}}^0\rangle=en_0$ that
couples to the electromagnetic gauge field $A_\mu$.  However, there is the ion
background with a positive charge to ensure the charge neutrality.
We take an isotropic elastic medium as the simplest model Lagrangian of ions,
\begin{equation}
n_0M[\dot{\vec{u}}^2-c_L^2(\vec{\nabla}\cdot\vec{u})^2-c_T^2(\vec{\nabla}\times\vec{u})^2]/2
-e A_\mu j_{\text{(bg)}}^\mu,
\end{equation}
where $M$ are the charge and the mass of the ions, $c_{L,T}$ are the longitudinal and the transverse phonon velocities, and
$ej_{\text{(bg)}}^\mu=-en_0(1-\vec{\nabla}\cdot\vec{u},\dot{\vec{u}})+O(u^2)$ is the
current density of the ions.  If we canonically normalize the
displacement field $\vec{u}$ so that the coefficient of
$\dot{\vec{u}}^2$ term becomes $1/2$, the coupling to the gauge
field is suppressed by the factor of $M^{-1/2}$ per $\vec{u}$ field.
Therefore, in the limit $M\rightarrow\infty$, the ion dynamics should be
completely decoupled from the rest of the system, while the charge density
$e\langle j_{\text{(bg)}}^0\rangle=-en_0$ still electrically
neutralizes the system.  In this decoupling limit, the background contribution is properly taken into account just by adding $-e A_\mu\langle j_{\text{(bg)}}^\mu\rangle=en_0A_t$ to the electron Lagrangian.

In general, Abelian charge densities can be consistently neutralized by a background~\cite{Kapusta}.  Including $-eA_\mu\langle j_{\text{(bg)}}^\mu\rangle$ does not explicitly break the gauge symmetry, since this term changes only by a surface term $\partial_\mu(-\chi \langle j_{\text{(bg)}}^\mu\rangle)$ under the gauge transformation $A_\mu\rightarrow A_\mu+e^{-1}\partial_\mu\chi$.

We now move on to non-Abelian charge densities.   In Ref.~\cite{Hama}, Hama and his collaborators proposed an extension of the above prescription to non-Abelian charge densities. Namely, they simply subtracted the background charge densities $2\text{tr}[eA_\mu\langle j_{\text{(bg)}}^\mu\rangle]$ from the Lagrangian. In the case of the ferromagnet, if we follow this proposed prescription, we add $em_3A_{\mu=t}^{I=3}$ ($\langle j_{\text{(bg)}}^\mu\rangle=-\delta^{\mu0}m_3T_3$) to the Lagrangian and find that Eq.~\eqref{ferrounitarygauge} is replaced by
\begin{eqnarray}
\mathcal{L}=-(\rho/2)e^2\vec{A}^a\cdot\vec{A}^a+\mathcal{L}_{\text{(gauge)}}.
\end{eqnarray}
This Lagrangian coincides with Eq.~\eqref{antiferrounitarygauge} with
$v\rightarrow\infty$.  In particular, it does not contain $m_3$ at all. Hence, there is no hope to reproduce the spectrum at $e=0$ in the $e\rightarrow0$ limit.  As explained above, one can observe a similar discontinuity in their original paper as well~\cite{Hama}.

We attribute this sick behavior to the incorrect treatment of the background degrees of freedom. The expectation value $\langle j_{\text{(bg)}}^\mu\rangle$ cannot transform under the symmetry transformation and the added term $2\text{tr}[eA_\mu\langle j_{\text{(bg)}}^\mu\rangle]$ {\it explicitly} breaks the
symmetry $A_\mu\rightarrow g A_\mu g^{-1}-ie^{-1}g\partial_\mu g^{-1}$, even the global one.  It is therefore mandatory to explicitly consider the background dynamics to recover the correct symmetry of the Lagrangian.

\paragraph{Two copies of ferromagnet.}---
This observation motivates us to theoretically consider two copies of ferromagnets, one with the magnetization $m_3$ and the other with $-m_3$.  We assume $G=\text{SO}(3)\times\text{SO}(3)$ symmetry, which is broken down to $H=\text{SO}(2)\times\text{SO}(2)$ by magnetizations.  In this case, we can safely gauge the diagonal $\text{SO}(3)$ symmetry thanks to the cancelation of the net magnetization.  Note that, due to the non-Abelian nature of the $\text{SO}(3)$ symmetry, gauging the vector part of $\text{SO}(3)\times\text{SO}(3)$ explicitly breaks the axial part of the global symmetry.

Denoting Nambu-Goldstone fields for the first (second) ferromagnet by $\pi^a$ ($\Pi^a$) ($a=1,2$), the linearized Lagrangian of the whole system reads
\begin{eqnarray}
\mathcal{L}&=&-(1/2)\rho(\vec{\nabla}\pi^a-e\vec{A}^a)^2\notag
-(1/2)\rho'(\vec{\nabla}\Pi^a-e\vec{A}^a)^2\notag\\
&&+m_3\epsilon_{ab}\left[(1/2)(\Pi^a\dot{\Pi}^b-\pi^a\dot{\pi}^b)+e\Pi^aA_t^b-e\pi^aA_t^b\right]\notag\\
&&+(1/2)(\vec{\nabla} A_t^I+\partial_t\vec{A}^I)^2-(c^2/2)(\vec{\nabla}\times \vec{A}^I)^2,\label{copy}
\end{eqnarray}

When $e=0$, the Lagrangian describes two type-B NGBs with $\omega_{B_1}=\rho k^2/m_3$ and $\omega_{B_2}=\rho'k^2/m_3$, and $3(d-1)$ photons with $\omega=ck$

For a finite coupling $e\neq0$, we use the unitary gauge to set $\pi^a+\Pi^a=0$. It is then straightforward~\cite{SM} to verify that the $(\vec{A}_L^a,\pi^a,A_t^a)$ sector describes two modes with
\begin{equation}
\omega_{L_{1,2}}=\sqrt{e^2(\rho+\rho')+\frac{(\rho+\rho')^2k^4}{(2m_3)^2}}\pm
\frac{(\rho-\rho')k^2}{2m_3},\label{L1}
\end{equation}
and $\vec{A}_T^I$ describes six transverse modes with
\begin{eqnarray}
\omega_T=\sqrt{e^2(\rho+\rho')+(vk)^2}.\label{transverse2}
\end{eqnarray}
In the limit of switching off the gauge coupling, $\omega_{L_{1,2}}$ smoothly go back to two type-B NGBs
$\omega_{B_{1,2}}$.  Therefore, the two type-B NGBs at $e=0$ are ``eaten'' to become the two
longitudinal gauge bosons at a finite coupling.  

As we discussed earlier, in order to maintain the unbroken spatial rotation, there must be the same number of gapless modes at $e=0$ as the number of massive spin-one boson at $e\neq0$.  A single ferromagnet alone had only one, and the background supplements the other one.

In order to implement the cancellation of the charge density for
general symmetry breaking $G/H$, we prepare a ``copy'' as above, but
we need to avoid copies of type-A NGBs.  Otherwise the end result has
unwanted extra gapless type-A NGBs.  Therefore, it is crucial that we
project $G/H$ down to a symplectic homogeneous space $G/U$ as
discussed by current authors in Ref.~\cite{WatanabeMurayama5}.
Fortunately, this projection is proven to be possible whenever the
symmetries can be gauged.  

\paragraph{Conclusion.}---
In this Letter, we clarified several issues regarding the EBH
mechanism in non-Lorentz-invariant systems.  There are two physically
distinct setups, with or without charge neutrality of matter currents.
We proved that, in the presence of finite non-Abelian matter charge
densities, the spatial rotation must always be broken in the weak
coupling regime.  We also showed that the naive subtraction of
background non-Abelian charge densities, which is proposed in Ref.~\cite{Hama},
explicitly breaks the symmetry of the system and results in
discontinuities of the spectrum as a function of the gauge coupling.  This sick behavior is cured by fully including the dynamics of the background.

It would be fascinating to see if these mechanisms can be realized in
real systems.  It is also an important future work to includes Chern-Simons terms.

\begin{acknowledgments}
H. W. is very grateful to Igor Shovkovy and Tomoya Hayata for helpful discussions.  
H. W. appreciates the financial support of the Honjo International
  Scholarship Foundation.  The work of H. M. was supported by the
  U.S. DOE under Contract DE-AC02-05CH11231, and by the NSF under
  grants PHY-1002399 and PHY-1316783.  H. M. was also supported by the
  JSPS Grant-in-Aid for Scientific Research (C) (No.~26400241),
  Scientific Research on Innovative Areas (No.~26105507), and by WPI,
  MEXT, Japan.
\end{acknowledgments}

\bibliography{references_Higgs}

\clearpage
\appendix

\onecolumngrid
\section{SUPPLEMENTAL MATERIAL\\for ``Englert--Brout--Higgs Mechanism in Nonrelativistic Systems"}

\subsection{The effective Lagrangian of gauged magnets}
Here we derive the effective Lagrangian of the gauged ferromagnet and antiferromagnet.  We assume $G=\text{SO}(3)$ and $H=\text{SO}(2)$, neglecting the time-reversal symmetry.  Let us first define the gauged Maurer-Cartan form $\Omega(\pi,A)$
via 
\begin{equation}
\Omega_\mu^IT_I\equiv-i U^{\dagger}(\partial_\mu+ie A_\mu)U,
\end{equation}
where $T_I=\sigma_I/2$ ($\sigma_{I=1,2,3}$ are Pauli matrices), $A_\mu\equiv A_\mu^IT_I$, and $U(\pi)\equiv e^{i\pi^a T_a}$ ($a,b=1,2$ are broken indices).  $(\pi^1,\pi^2)$ is a local coordinate of the coset space $G/H=S^2$. For example, one can use $\pi^1=\theta\sin\phi$ and $\pi^2=-\theta\cos\phi$, where $(\theta,\phi)$ is the spherical coordinate of $S^2$.

A local symmetry transformation $g\in G$ acts on NG fields
$\pi^a$ and the gauge field $A_\mu$ as
\begin{eqnarray}
gU(\pi)&=&U(\pi')h,\quad h\equiv e^{i\theta(g,\pi)T_3}\in H,\\
A_\mu'&=&gA_\mu g^{-1}-ie^{-1}g \partial_\mu g^{-1},
\end{eqnarray}
so that the Maurer-Cartan form transforms nicely,
\begin{eqnarray}
(\Omega_\mu^a)'=R(\theta)^a_{\,\,b}\Omega_\mu^b,\quad (\Omega_\mu^z)'=\Omega_\mu^z-\partial_\mu \theta,\label{MCtransformation}
\end{eqnarray}
where $R(\theta)$ is the rotation matrix around the $z$ axis by an angle $\theta(g,\pi)$.

The most general effective Lagrangian to the quadratic order in derivatives is given by
\begin{equation}
\mathcal{L}_{(\text{matter})}=\frac{\rho}{2v^2}(\Omega_{t}^{a})^2-m\Omega_{\mu=t}^{I=3}-\frac{\rho}{2}(\Omega_{i}^{a})^2.
\end{equation}
Here, the ferromagnet corresponds to $v\rightarrow\infty$ and $m>0$, while the antiferromagnet to $0<v<\infty$ and $m=0$.
Using the transformation rule in Eq.~\eqref{MCtransformation}, one can easily check that $\mathcal{L}_{(\text{matter})}$ is invariant up to a surface term under a {\it local} $G=\text{SO}(3)$ transformation. 

At $e=0$, the Lagrangian reduces to
\begin{equation}
\mathcal{L}_{(\text{matter})}=\frac{\rho}{2v^2}\partial_t\vec{n}\cdot\partial_t\vec{n}+m_3\frac{n_2\dot{n}_1-n_1\dot{n}_2}{1+n_3}-\frac{\rho}{2}\partial_i\vec{n}\cdot\partial_i\vec{n},
\end{equation}
Here, $\vec{n}$ is the unit vector and the second term can be written as $m_3(-1+\cos\theta)\partial_t\phi$ in terms of the spherical coordinate.

On the other hand, for a finite $e\neq0$, one may take the unitary gauge $\pi^a=0$.  Then we have
\begin{equation}
\mathcal{L}_{(\text{matter})}=\frac{\rho e^2}{2v^2}(A_{t}^{a})^2-em_3 A_{\mu=t}^{I=3}-\frac{\rho e^2}{2}(A_{i}^{a})^2.
\end{equation}

\subsection{The spectrum of ferromagnet at a finite coupling $e\neq0$}
The total Lagrangian for the gauged ferromagnet in the unitary gauge reads
\begin{equation}
\mathcal{L}=-em_3 A_{\mu=t}^{I=3}-\frac{\rho e^2}{2}(A_{i}^{a})^2+\text{tr}F_{ti}F_{ti}-\frac{c^2}{2}\text{tr}F_{ij}F_{ij}.
\end{equation}

Among homogenous field configurations, the minimal energy is achieve by
\begin{equation}
\langle A_{\mu=t}^{I=3}\rangle=\sqrt{\rho},\quad \langle A_{\mu=z}^{I=1}\rangle=-\sqrt{\frac{m_3}{e\sqrt{\rho}}}.
\end{equation}
We expand the Lagrangian to the quadratic order in fluctuations $a_\mu^I\equiv A_\mu^I-\langle A_\mu^I\rangle$. The result is particularly simple when we set the momentum along the $z$-axis $\vec{k}=k\hat{z}$.  In that case, we have
\begin{eqnarray}
\frac{1}{2}\sum_{i=x,y}\begin{pmatrix}
a_i^1&a_i^2&a_i^3
\end{pmatrix}^*
G_i
\begin{pmatrix}
a_i^1\\a_i^2\\a_i^3
\end{pmatrix}
+\frac{1}{2}\begin{pmatrix}
a_z^1&a_z^2&a_z^3&a_t^1&a_t^2&a_t^3
\end{pmatrix}^*
G_z
\begin{pmatrix}
a_z^1\\a_z^2\\a_z^3\\a_t^1\\a_t^2\\a_t^3
\end{pmatrix},
\end{eqnarray}
where
\begin{equation}
G_x=G_y=\begin{pmatrix}
\omega^2-c^2k^2&2e\sqrt{\rho}i\omega&0\\
-2e\sqrt{\rho}i\omega&\omega^2-c^2k^2-\frac{em_3c^2}{\sqrt{\rho}}&-\frac{2c^2\sqrt{em}_3ik}{\rho^{1/4}}\\
0&\frac{2c^2\sqrt{em_3}ik}{\rho^{1/4}}&\omega^2-c^2k^2-\frac{em_3c^2}{\sqrt{\rho}}
\end{pmatrix}
\end{equation}
and
\begin{equation}
G_z=\begin{pmatrix}
\omega^2&2e\sqrt{\rho}i\omega&0&k\omega&e\sqrt{\rho}ik&-2e^{3/2}\rho^{1/4}\sqrt{m_3}\\
-2e\sqrt{\rho}i\omega&\omega^2&0&-e\sqrt{\rho}ik&k\omega&\frac{\sqrt{em_3}i\omega}{\rho^{1/4}}\\
0&0&\omega^2-\frac{c^2k^2}{\xi}&e^{3/2}\rho^{1/4}\sqrt{m_3}&-\frac{\sqrt{em_3}i\omega}{\rho^{1/4}}&k\omega-\frac{k\omega}{\xi}\\
k\omega&e\sqrt{\rho}ik&e^{3/2}\rho^{1/4}\sqrt{m_3}&k^2&0&0\\
-e\sqrt{\rho}ik&k\omega&\frac{\sqrt{em_3}i\omega}{\rho^{1/4}}&0&k^2+\frac{em_3}{\sqrt{\rho}}&\frac{2\sqrt{em_3}ik}{\rho^{1/4}}\\
-2e^{3/2}\rho^{1/4}\sqrt{m_3}&-\frac{\sqrt{em_3}i\omega}{\rho^{1/4}}&k\omega-\frac{k\omega}{\xi}&0&-\frac{2\sqrt{em_3}ik}{\rho^{1/4}}&k^2+\frac{em_3}{\sqrt{\rho}}-\frac{\omega^2}{c^2\xi}
\end{pmatrix}.
\end{equation}
Here, we added the gauge fixing term $(2\xi)^{-1}(\partial_\mu A_{I=3}^\mu)^2$ to fix the unfixed gauge redundancy.  The low-energy spectrum shown in the main text is obtained by solving $\text{det}G=0$.  One can of course set momentum along arbitrary directions and do the same thing.

\end{document}